\documentclass[notitlepage,nofootinbib,preprintnumbers,amssymb,superscriptaddress]{revtex4-1}
\usepackage{amsfonts,amssymb,amsmath,graphicx,color,bm}
\usepackage[linktocpage=true]{hyperref}
\hypersetup{
colorlinks=true,
citecolor=red,
linkcolor=blue,
urlcolor=blue,
}

\begin{document}

\title{On the gravitational waves coupled with electromagnetic waves}

\author{A.N. Morozov}
\email[E-mail:]{amor59@mail.ru, amor@bmstu.ru}
\affiliation{Department of Physics, Bauman Moscow State Technical University,
  Moscow, 105005, Russia}

\author{V.I. Pustovoit}
\email[E-mail:]{vladpustovoit@gmail.com}
\affiliation{Department of Physics, Bauman Moscow State Technical University,
  Moscow, 105005, Russia}
\affiliation{Scientific and Technological Centre of Unique Instrumentation of the Russian Academy of Sciences, Moscow, 117342, Russia}

\author{I.V. Fomin}
\email[E-mail:]{ingvor@inbox.ru}
\affiliation{Department of Physics, Bauman Moscow State Technical University,
  Moscow, 105005, Russia}

%\preprint{1234}

\begin{abstract}
A description is made of the process of excitation of coupled longitudinal-transverse gravitational waves during the propagation of a strong electromagnetic wave in a vacuum and when a standing electromagnetic wave exists in the Fabry-Perot resonator. It is shown that such waves lead to the appearance of transverse gravitational waves in empty space. It was established that two standing high-frequency electromagnetic waves in a Fabry-Perot resonator with close frequencies cause the appearance of a low-frequency transverse gravitational wave in empty space.
\end{abstract}

\maketitle

\section{Introduction}

In the framework of predictions of the General Relativity~\cite{Einstein},
it is believed that weak gravitational waves in empty space are transverse only~\cite{Eddington,Landau,Weber,Maggiore:2018sht}.
It is precisely such gravitational waves that arose as a result of the black holes and neutron stars merger that were discovered in the LIGO and Virgo experiments~\cite{Aasi:2013wya,Abbott:2016blz,Abbott:2016nmj,Monitor:2017mdv}
by using the interference method firstly proposed in~\cite{Gertsenshtein:1962kfm}.
It should also be noted that in the case of extended metric theories of gravity, gravitational waves have six types
of polarization, namely, two tensor, two vector and two scalar
ones~\cite{Liang:2017ahj,Hou:2017bqj,Hyun:2018pgn,Wagle:2019mdq,Shankaranarayanan:2019yjx}.
Vector and scalar modes were not directly observed, nevertheless, the possibility of their observation determines
the method of additional experimental verification of the extensions of General Relativity.
For verification of possible modifications of Einstein gravity by detecting these additional modes of gravitational waves from astrophysical sources and in the study of stochastic gravitational-wave background or confirming their absence in future experiments together with a further analysis of already discovered tensor modes the interference method can be used as well~\cite{Gertsenshtein:1962kfm,Nishizawa:2009bf,Philippoz:2017ywb,Abbott:2018utx,Takeda:2018uai}.

A lot of works have been devoted to studying the properties of transverse gravitational waves
(tensor modes) and the developing methods for their emission and registration
(see, for example~\cite{Grishchuk:1973qz,Denisov,Nikishov:2010zz};
for a review, see~\cite{Hough:2005pf,Chen:2016isk,Pustovoit:2016zyt,Rudenko:2017asr}).
Moreover, in all these works, it is believed that for the case of a weak gravitational wave,
only transverse waves with two polarizations in the framework of GR can exist~\cite{Grishchuk_1977}.
However, this statement is strictly proved only for the case of the propagation of a gravitational wave in empty space~\cite{Eddington,Landau,Weber,Maggiore:2018sht}.

Weak gravitational waves are considered  based on the linearized Einstein gravity theory as small perturbations of Minkowski space-time~\cite{Eddington,Landau,Weber,Maggiore:2018sht}
\begin{equation}
\label{Minkowski}
g_{\mu\nu}=\eta_{\mu\nu}+h_{\mu\nu},~~~|h_{\mu\nu}|\ll 1,
\end{equation}
where $\eta_{\mu\nu}$ is the metric tensor of Minkowski space-time
with nonzero components  $\eta_{00}=1$, $\eta_{11}=\eta_{22}=\eta_{33}=-1$
and $\mu,\nu=0,1,2,3$.

In this case, the Einstein equations, taking into account the  harmonic gauge, can be written as follows \cite{Eddington,Landau,Weber,Maggiore:2018sht}
\begin{eqnarray}
\label{GW}
\Box h_{\mu\nu}=\Delta h_{\mu\nu}-\frac{1}{c^{2}}\frac{\partial^{2}h_{\mu\nu}}{\partial t^{2}}=
\frac{16\pi G}{c^{4}}\left(T_{\mu\nu}-\frac{1}{2}\eta_{\mu\nu}T\right),
\end{eqnarray}
where $G$ is the gravitational constant, $c$ is the velocity of light in vacuum and $T_{\mu\nu}$ is the energy-momentum tensor.

For gravitational waves propagating in direction $x^{1}=x$, non-zero components of tensor $h_{\mu\nu}$ determine three possible types \cite{Eddington}:
\begin{itemize}
\item  $h_{22}$,  $h_{23}$, $h_{33}$ -- transverse--transverse ($TT$),
\item $h_{12}$,  $h_{13}$, $h_{20}$, $h_{30}$ -- longitudinal--transverse ($LT$),
\item $h_{11}$, $h_{10}$, $h_{00}$ -- longitudinal--longitudinal ($LL$).
\end{itemize}

However, in empty space ($T_{\mu\nu}=0$), $LT$ and $LL$--waves do not lead to a deviation from the flat Minkowski space-time and can be eliminated by additional coordinate transformations~\cite{Eddington}.
Also, one has three possible polarizations for $TT$-waves, namely $h_{23}$, $h_{22}-h_{33}$ and $h_{22}+h_{33}$.
However, equations (\ref{GW}) in empty space lead to condition $h_{22}+h_{33}=0$, and, thus,
transverse-traceless gravitational waves with only two polarizations $h_{23}$ and $h_{22}-h_{33}$
can propagate, which corresponds to so called transverse-traceless gauge~\cite{Maggiore:2018sht}.

Nevertheless, one can consider the possibility of the existence of coupled states of electromagnetic
and gravitational waves for which $LT$ and $LL$--types of the metric perturbations are non-zero and
$h_{22}+h_{33}\neq0$ on the basis of Einstein-Maxwell equations written in a general form.

The propagation of $TT$--waves with polarization $h_{22}+h_{33}$ in the field of an electromagnetic wave
was considered earlier in the paper~\cite{Eddington1}.
In~\cite{Gertsenshtein1962,Zel'dovich1974,Gerlach:1974zz,Zeldovich1983,Raffelt:1987im,Pustovoit:1981za,Fargion:1995mm,Marklund:1999sp,
Dolgov:2012be,Kolosnitsyn:2015zua,Dolgov:2017bpj,Ejlli:2018hke}, the possibility of generating
gravitational waves during the propagation of a strong electromagnetic wave in a constant magnetic field was
considered, and in~\cite{Zel'dovich1974}, the possibility of resonant excitation of coupled longitudinal--transverse
gravitational waves arising from the propagation of a strong electromagnetic wave in vacuum, which with subsequent
 interaction lead to the generation of the third harmonic of the electromagnetic wave. Such resonant excitation of the
 indicated gravitational wave is a consequence of the equality of the propagation velocities of electromagnetic and
 gravitational waves, and therefore it is possible to fulfill the synchronism conditions at large distances.
 As shown by estimates of recent direct observations of gravitational waves, synchronism is carried
 out up to the 15th sign~\cite{Monitor:2017mdv}. It is physically clear that such a connected
 longitudinal-transverse gravitational wave should also arise for the case of a strong standing electromagnetic wave.
Also, the existence of the additional modes in the framework of General Relativity as evanescent gravitational waves
similar to ones in acoustics and optics is discussed in~\cite{Golat:2019aap}.
%Also, we note, that the gravitational fields induced by electromagnetic waves are considered in \cite{Tolman:1931zza,Aichelburg:1970dh,Scully:1979xe,Ratzel:2015nqf}.

The purpose of this paper is the analysis of gravitational waves coupled with an electromagnetic wave, otherwise,
the gravi-electromagnetic ones and their further transformation into $TT$--gravitational waves in vacuum after
absorption of an electromagnetic wave.

The article is organized as follows:  in section \ref{sect2} we describe the coupled gravitational waves induced
by electromagnetic field on the basis of Einstein-Maxwell equations. In section \ref{sect3} we consider such a
gravitational waves coupled with a propagating strong electromagnetic wave and their transformation during
the absorption of an electromagnetic wave. Section \ref{sect4} discusses the generation of gravitational waves
through a standing electromagnetic wave. In section \ref{sect5}, we calculate the characteristics of gravitational
waves induced by two standing electromagnetic waves, which makes it possible to generate low-frequency
gravitational waves at the difference frequency. In conclusion, the results of this study are discussed.

\section{Description of a coupled gravitational wave}\label{sect2}

To describe a coupled plane gravitational wave propagating in the direction $x^{1}=x$, we will use the
Einstein-Maxwell equations describing the gravitational field corresponding to the electromagnetic one
\begin{equation}
\label{EEQ}
R_{\mu\nu}-\frac{1}{2}g_{\mu\nu}R=\frac{8\pi G}{c^{4}}T_{\mu\nu},
\end{equation}
Also, we note that for any electromagnetic field, the scalar curvature $R=0$~\cite{Landau}.

The components of the energy-momentum tensor of electromagnetic field in vacuum can be written as~\cite{Landau}
\begin{eqnarray}
\label{Energy}
&&T_{00}=\frac{1}{8\pi}\left(E^{2}+H^{2}\right),\\
\label{Pointing}
&&T_{0k}=T_{k0}=-\frac{1}{4\pi}[{\bf E}\times{\bf H}]_{k}=
-\frac{1}{4\pi}\varepsilon^{kij}E_{i}H_{j},\\
\label{Maxvell}
&&T_{ij}=-\frac{1}{4\pi}\left[E_{i}E_{j}+
H_{i}H_{j}-\frac{1}{2}\delta_{ij}\left(E^{2}+H^{2}\right)\right],
\end{eqnarray}
where $i,j,k=1,2,3$, $\varepsilon^{kij}$ is the Levi-Civita symbol and $\delta_{ij}$ is the Kronecker one.

For a plane electromagnetic wave propagating in the direction $x^{1}=x$ in the general case
(including the case of propagating and standing electromagnetic waves), the components of the energy-momentum
 tensor have the following form: $T_{00}=T_{11}$, $T_{01}=T_{10}$ and $T_{22}=-T_{33}$
 (for propagating electromagnetic wave one has $T_{22}=-T_{33}=0$). The remaining components of
 the energy-momentum tensor $T_{\mu\nu}$ are equal to zero.

Also, the momentum energy tensor must satisfy the conservation laws
\begin{equation}
\label{CL}
\frac{\partial T^{\mu\nu}}{\partial x^{\nu}}=0.
\end{equation}

The Ricci tensor can be written as
\begin{equation}
\label{Ricci}
R_{\mu\nu}=g^{\sigma\rho}R_{\sigma\mu\rho\nu},
\end{equation}
where the Riemann tensor for the case of a weak gravitational field is
\begin{equation}
\label{Riemann}
R_{\sigma\mu\rho\nu}=\frac{1}{2}\left(\frac{\partial^{2} h_{\sigma\nu}}{\partial x^{\mu}\partial x^{\rho}}+
\frac{\partial^{2} h_{\mu\rho}}{\partial x^{\sigma}\partial x^{\nu}}-\frac{\partial^{2} h_{\sigma\rho}}{\partial x^{\mu}\partial x^{\nu}}-\frac{\partial^{2} h_{\mu\nu}}{\partial x^{\sigma}\partial x^{\rho}}\right).
\end{equation}

Thus, the Einstein-Maxwell equations (\ref{EEQ}) for the case of the propagation of a
plane gravitational wave in the direction $x^{1}=x$ have the following form
\begin{eqnarray}
\label{E1}
&&-\frac{\partial^{2} h_{11}}{c^{2}\partial t^{2}}+2\frac{\partial^{2} h_{01}}{c\partial t\partial x}-
\frac{\partial^{2} h_{00}}{\partial x^{2}}+\frac{\partial^{2} h_{22}}{c^{2}\partial t^{2}}+
\frac{\partial^{2} h_{33}}{c^{2}\partial t^{2}}=\frac{16\pi G}{c^{4}}T_{00},\\
\label{E2}
&&\frac{\partial^{2} h_{11}}{c^{2}\partial t^{2}}-2\frac{\partial^{2} h_{01}}{c\partial t\partial x}+
\frac{\partial^{2} h_{00}}{\partial x^{2}}+\frac{\partial^{2} h_{22}}{\partial x^{2}}+
\frac{\partial^{2} h_{33}}{\partial x^{2}}=\frac{16\pi G}{c^{4}}T_{11},\\
\label{E3}
&&\frac{\partial^{2} h_{22}}{c\partial t\partial x}+\frac{\partial^{2} h_{33}}{c\partial t\partial x}
=\frac{16\pi G}{c^{4}}T_{01},\\
\label{E4}
&&\frac{\partial^{2} h_{22}}{\partial x^{2}}-\frac{\partial^{2} h_{22}}{c^{2}\partial t^{2}}
=\frac{16\pi G}{c^{4}}T_{22},\\
\label{E5}
&&\frac{\partial^{2} h_{33}}{\partial x^{2}}-\frac{\partial^{2} h_{33}}{c^{2}\partial t^{2}}
=\frac{16\pi G}{c^{4}}T_{33},
\end{eqnarray}
where the condition $h_{01}=h_{10}$ was used.

Addition of (\ref{E1}) and (\ref{E2}) gives
\begin{equation}
\label{E6}
\frac{\partial^{2} \tilde{h}}{c^{2}\partial t^{2}}+\frac{\partial^{2} \tilde{h}}{\partial x^{2}}=\frac{32\pi G}{c^{4}}T_{00},
\end{equation}
and addition (\ref{E4}) and (\ref{E5}) leads to
\begin{equation}
\label{E7}
\frac{\partial^{2} \tilde{h}}{\partial x^{2}}-\frac{\partial^{2} \tilde{h}}{c^{2}\partial t^{2}}=0,
\end{equation}
where
\begin{equation}
\label{E8}
\tilde{h}= h_{22}+h_{33}
\end{equation}
Then, taking into account (\ref{E6})--(\ref{E8}), the system of equations (\ref{E1})--(\ref{E5}) can be written as
\begin{eqnarray}
\label{E9}
&&\frac{\partial^{2} h_{11}}{c^{2}\partial t^{2}}-2\frac{\partial^{2} h_{01}}{c\partial t\partial x}+
\frac{\partial^{2} h_{00}}{\partial x^{2}}=0,\\
\label{E10}
&&\frac{\partial^{2} \tilde{h}}{c^{2}\partial t^{2}}+\frac{\partial^{2} \tilde{h}}{\partial x^{2}}=\frac{32\pi G}{c^{4}}T_{00},\\
\label{E11}
&&\frac{\partial^{2} \tilde{h}}{\partial x^{2}}-\frac{\partial^{2} \tilde{h}}{c^{2}\partial t^{2}}=0,\\
\label{E12}
&&\frac{\partial^{2} \tilde{h}}{c\partial t\partial x}=\frac{16\pi G}{c^{4}}T_{01},\\
\label{E13}
&&\frac{\partial^{2} h_{22}}{\partial x^{2}}-\frac{\partial^{2} h_{22}}{c^{2}\partial t^{2}}
=\frac{16\pi G}{c^{4}}T_{22},\\
\label{E14}
&&\frac{\partial^{2} h_{33}}{\partial x^{2}}-\frac{\partial^{2} h_{33}}{c^{2}\partial t^{2}}
=\frac{16\pi G}{c^{4}}T_{33}.
\end{eqnarray}

It follows from equation (\ref{E9}) that the determination of the components of tensor
$h_{00}$, $h_{01}=h_{10}$ and $h_{11}$ from the system of equations (\ref{E9})--(\ref{E14})
is impossible without additional conditions.
Such an additional conditions can be obtained from the harmonic gauge, which has the following form
\begin{equation}
\label{HC}
\frac{\partial}{\partial x^{\nu}}\left(h^{\nu}_{\mu}-\frac{1}{2}\delta^{\nu}_{\mu}h\right)=
\frac{\partial}{\partial x^{\nu}}\left(\eta^{\nu\sigma}h_{\sigma\mu}-\frac{1}{2}\delta^{\nu}_{\mu}h\right)=0,
\end{equation}
where
\begin{equation}
\label{HCA}
h=h_{00}-h_{11}-h_{22}-h_{33}=-(h_{22}+h_{33})=-\tilde{h},
\end{equation}
for $h_{00}=h_{11}$.

For the case under consideration, equation (\ref{HC}) for $\nu=0$ and $\nu=1$ gives
\begin{eqnarray}
\label{HCA1}
&&\frac{\partial}{c\partial t}\left(h_{00}+\frac{1}{2}\tilde{h}\right)-\frac{\partial h_{01}}{\partial x}=0,\\
\label{HCA2}
&&-\frac{\partial}{\partial x}\left(h_{11}-\frac{1}{2}\tilde{h}\right)+\frac{\partial h_{01}}{c\partial t}=0.
\end{eqnarray}

The system of equations (\ref{E9})--(\ref{E14}) together with (\ref{HC}) allows us to calculate the tensor components
$h_{00}$, $h_{01}=h_{10}$, $h_{11}$, $h_{22}$, $h_{33}$ and the sum $\tilde{h}= h_{22}+h_{33}$.
It should be noted, that the components $h_{22}$ and $h_{33}$ due to equation (\ref{E7}) are determined up to
a satisfactory harmonic gauge of function $\tilde{h}$.

Thus, the non-zero components of gravitational waves coupled with an electromagnetic wave in this case are
$LL$--waves and $TT$--waves with polarization $h_{22}+h_{33}\neq0$. Therefore, one can consider such a
gravitational waves as longitudinal-transverse ones.

\section{Propagation of a strong electromagnetic wave in a vacuum}\label{sect3}

Let a plane harmonic electromagnetic wave be emitted from a point $x=0$ in the direction of the axis $x^{1}=x$.
Such a wave can be described by following expressions for the electric and magnetic fields
\begin{eqnarray}
\label{EF}
&&E_{y}=E_{0}\cos\left(\omega\left(t-\frac{x}{c}\right)\right),\\
\label{MF}
&&H_{z}=H_{0}\cos\left(\omega\left(t-\frac{x}{c}\right)\right),
\end{eqnarray}
where $E_{0}=H_{0}$.

For the case, from expressions (\ref{Energy})--(\ref{Maxvell}) one has following non-zero components
\begin{eqnarray}
\label{TEF1}
&&T_{00}=T_{11}=\frac{1}{8\pi}\left(E^{2}+H^{2}\right),\\
\label{TEF2}
&&T_{01}=T_{10}=-\frac{1}{4\pi}E_{y}H_{z}.
\end{eqnarray}

After substituting (\ref{EF})--(\ref{MF}) into  (\ref{TEF1})--(\ref{TEF2}) we obtain
\begin{eqnarray}
\label{TEF3}
&&T_{00}=T_{11}=\frac{E^{2}_{0}}{8\pi}\left[1+\cos\left(2\omega\left(t-\frac{x}{c}\right)\right)\right],\\
\label{TEF4}
&&T_{01}=T_{10}=-\frac{E^{2}_{0}}{8\pi}\left[1+\cos\left(2\omega\left(t-\frac{x}{c}\right)\right)\right],
\end{eqnarray}
therefore, one can consider the components of the energy-momentum tensor as sums of variable and constant parts.

%Also, one can consider the non-zero components of the energy-momentum tensor as
%\begin{eqnarray}
%\label{TEF5}
%&&T_{00}=T_{11}=\frac{E^{2}_{0}}{8\pi}\cos\left(2\omega\left(t-\frac{x}{c}\right)\right)+\frac{E^{2}_{0}}{8\pi},\\
%\label{TEF6}
%&&T_{01}=T_{10}=-\frac{E^{2}_{0}}{8\pi}\cos\left(2\omega\left(t-\frac{x}{c}\right)\right)-\frac{E^{2}_{0}}{8\pi}.
%\end{eqnarray}

Such an energy-momentum tensor satisfies the conservation laws (\ref{CL}) for a weak gravitational field
\begin{eqnarray}
\label{TEF1CL}
&&\mu=0,~~~~~\frac{\partial T^{00}}{\partial x^{0}}+\frac{\partial T^{01}}{\partial x^{1}}
=\frac{\partial T_{00}}{c\partial t}-\frac{\partial T_{01}}{\partial x}=0,\\
\label{TEF2CL}
&&\mu=1,~~~~~\frac{\partial T^{10}}{\partial x^{0}}+\frac{\partial T^{11}}{\partial x^{1}}
=-\frac{\partial T_{10}}{c\partial t}+\frac{\partial T_{11}}{\partial x}=0.
\end{eqnarray}

The general solutions of the equations (\ref{E9})--(\ref{E14}) with the energy-momentum tensor (\ref{TEF3})--(\ref{TEF4})
for the components $h_{22}+h_{33}$ can be noted as
\begin{equation}
\label{GENSE}
h_{22}+h_{33}=-\frac{GE^{2}_{0}}{2c^{2}\omega^{2}}\cos\left(2\omega\left(t-\frac{x}{c}\right)\right)+
\frac{GE^{2}_{0}}{c^{2}}\left(t-\frac{x}{c}\right)^{2}+c_{1}x+c_{2}t+c_{3},
\end{equation}
where $c_{1}$, $c_{2}$ and $c_{3}$ are the constants of integration.

This solution contains the periodic and aperiodic parts that are induced by the variable and constant parts of the energy-momentum tensor, respectively.

Also, we note that these solutions were obtained earlier in Eddington's work \cite{Eddington1}. Following this work, we consider $c_{1}=c_{2}=0$ (one can also use the condition $c_{3}=0$ as well) and write the aperiodic part of the solution as
\begin{equation}
\label{APERIODIC}
\tilde{h}=h_{22}+h_{33}=\frac{GE^{2}_{0}}{c^{2}}\left(t-\frac{x}{c}\right)^{2}.
\end{equation}

By a similar way, taking into account (\ref{HCA1})--(\ref{HCA2}), one can obtain
the aperiodic parts of the solutions for the other components
\begin{eqnarray}
\label{NW1}
&&h_{00}=h_{11}=-\frac{GE^{2}_{0}}{c^{2}}t\left(t-\frac{x}{c}\right),\\
\label{NW2}
&&h_{01}=h_{10}=\frac{GE^{2}_{0}}{c^{2}}t\left(t-\frac{x}{c}\right).
\end{eqnarray}

The aperiodic part (\ref{APERIODIC}) indicate that field is undergoing a cumulative change.
Such a part of wave do actually carry away energy from the source and it is non negligible quantity
in the approximation of small field (\ref{Minkowski}).
As shown in \cite{Eddington1}, the gravitational field coupled with the electromagnetic wave can be associated
with the periodic part of the solution (\ref{GENSE}) only.

Therefore, we will consider the periodic solutions of the Einstein-Maxwell equations (\ref{E1})--(\ref{E5})
which associated with the variable part of the energy-momentum tensor for the description of such a gravitational waves.

Thus, one can write the  gravitational wave solutions of equations (\ref{E9})--(\ref{E14}) and (\ref{HCA1})--(\ref{HCA2})
for the energy-momentum tensor with non-zero components  (\ref{TEF3})--(\ref{TEF4}) as follows
\begin{eqnarray}
\label{WS1}
&&h_{00}=h_{11}=-\frac{GE^{2}_{0}}{2c^{3}\omega}x\sin\left(2\omega\left(t-\frac{x}{c}\right)\right),\\
\label{WS2}
&&h_{01}=h_{10}=\frac{GE^{2}_{0}}{2c^{3}\omega}x\sin\left(2\omega\left(t-\frac{x}{c}\right)\right),\\
\label{WS3}
&&\tilde{h}=h_{22}+h_{33}=-\frac{GE^{2}_{0}}{2c^{2}\omega^{2}}\cos\left(2\omega\left(t-\frac{x}{c}\right)\right).
\end{eqnarray}

From these formulas, we can write expressions for the tensor components $h_{22}$ and $h_{33}$ in the following form
\begin{eqnarray}
\label{WS4}
&&h_{22}=-\frac{\alpha GE^{2}_{0}}{2c^{2}\omega^{2}}\cos\left(2\omega\left(t-\frac{x}{c}\right)\right),\\
\label{WS5}
&&h_{33}=-\frac{\beta GE^{2}_{0}}{2c^{2}\omega^{2}}\cos\left(2\omega\left(t-\frac{x}{c}\right)\right),
\end{eqnarray}
where $\alpha$ and $\beta$ are arbitrary constants satisfying the condition $\alpha+\beta=1$.
From the symmetry and equality of axes $x^{2}=y$ and $x^{3}=z$ one can define these constants as $\alpha=\beta=1/2$.

The gravitational wave energy flux density takes the form
\begin{equation}
\label{EnFlux}
ct^{01}=-\frac{c^{5}}{32\pi G}\left(\frac{\partial h_{00}}{\partial x^{0}}\frac{\partial h_{00}}{\partial x^{1}}-
\frac{\partial h_{10}}{\partial x^{0}}\frac{\partial h_{01}}{\partial x^{1}}-
\frac{\partial h_{01}}{\partial x^{0}}\frac{\partial h_{10}}{\partial x^{1}}+\frac{\partial h_{11}}{\partial x^{0}}\frac{\partial h_{11}}{\partial x^{1}}+\frac{\partial h_{22}}{\partial x^{0}}\frac{\partial h_{22}}{\partial x^{1}}+\frac{\partial h_{33}}{\partial x^{0}}\frac{\partial h_{33}}{\partial x^{1}}\right).
\end{equation}

Substitution of expressions (\ref{WS1})--(\ref{WS5}) into (\ref{EnFlux}) gives
\begin{equation}
\label{EnFlux1}
ct^{01}=\frac{GE^{4}_{0}}{32\pi c\omega^{2}}\left(\alpha^{2}+\beta^{2}\right)\sin\left(2\omega\left(t-\frac{x}{c}\right)\right)=
\frac{GE^{4}_{0}}{64\pi c\omega^{2}}\sin\left(2\omega\left(t-\frac{x}{c}\right)\right).
\end{equation}
It should be noted here that the first four terms in parentheses in expression (\ref{EnFlux}) add up to zero in total.

We also note that when substituting solutions (\ref{WS1}) and (\ref{WS2}) into the equation for a strong electromagnetic wave, the generation of a weak electromagnetic wave at the third harmonic occurs, which was predicted in~\cite{Pustovoit:1981za}. This, along with the presence of energy flux density, indicates the existence of longitudinal-transverse electromagnetic waves associated with the electromagnetic wave.

A separate task is to solve the problem of docking the boundary conditions at the surface of the area occupied by a coupled gravitational waves and an empty space in which this type of gravitational waves cannot exist.
Let a strong electromagnetic wave be completely absorbed on the surface with a coordinate $x=L$. Assuming that at this boundary the condition of conservation of the energy flux density of the gravitational wave is fulfilled, one can find the tensor components $h_{22}=-h_{33}$ describing the gravitational wave in empty space
\begin{equation}
\label{WS6}
h_{22}=-h_{33}=\pm\frac{GE^{2}_{0}\sqrt{\alpha^{2}+\beta^{2}}}{2\sqrt{2}c^{2}\omega^{2}}
\cos\left(2\omega\left(t-\frac{x}{c}\right)\right),
\end{equation}
and for $\alpha=\beta=1/2$ one has
\begin{equation}
\label{WS7}
h_{22}=-h_{33}=\pm\frac{GE^{2}_{0}}{4c^{2}\omega^{2}}
\cos\left(2\omega\left(t-\frac{x}{c}\right)\right).
\end{equation}
In this case it is not possible to determine the sign in these formulas and, thus, it can be either positive or negative.

On the basis of the above results, it can be concluded that an electromagnetic wave propagating in a vacuum can generate a transverse gravitational wave with the polarization $h_{22}-h_{33}$ in empty space when it is absorbed.

\section{Generating of a gravitational wave by a standing electromagnetic wave}\label{sect4}
In the paper \cite{Morozov2020}, a description was made of the excitation of a coupled longitudinal transverse wave by a standing electromagnetic wave in unlimited space. Let a plane standing electromagnetic wave be in the Fabry-Perot resonator between the mirrors located at the points of the axis $x=-L/2 $ and $x=L/2$, where $L$ is the length of the resonator. Then we can write the expressions for the electric and magnetic fields in this wave
\begin{eqnarray}
\label{EFS}
&&E_{y}=E_{0}\cos\left(\omega\left(t-\frac{x}{c}\right)\right)+E_{0}\cos\left(\omega\left(t+\frac{x}{c}\right)\right),\\
\label{MFS}
&&H_{z}=H_{0}\cos\left(\omega\left(t-\frac{x}{c}\right)\right)-H_{0}\cos\left(\omega\left(t+\frac{x}{c}\right)\right),
\end{eqnarray}
where $E_{0}=H_{0}$.

On the basis of equations (\ref{Energy})--(\ref{Maxvell}), we write the expressions for non-zero components of the energy-momentum tensor as
\begin{eqnarray}
\label{TEF1S}
&&T_{00}=T_{11}=\frac{E^{2}_{0}}{8\pi}\left[2+\cos\left(2\omega\left(t-\frac{x}{c}\right)\right)+
\cos\left(2\omega\left(t+\frac{x}{c}\right)\right)\right],\\
\label{TEF2S}
&&T_{01}=T_{10}=-\frac{E^{2}_{0}}{8\pi}\left[\cos\left(2\omega\left(t-\frac{x}{c}\right)\right)-
\cos\left(2\omega\left(t+\frac{x}{c}\right)\right)\right],\\
\label{TEF3S}
&&T_{22}=-T_{33}=-\frac{E^{2}_{0}}{2\pi}\left[\cos\left(2\omega\left(t-\frac{x}{c}\right)\right)
\cos\left(2\omega\left(t+\frac{x}{c}\right)\right)\right].
\end{eqnarray}
For components (\ref{TEF1S}) and (\ref{TEF2S}), conservation laws (\ref{TEF1CL}) and (\ref{TEF2CL}) are satisfied, and for expression (\ref{TEF3S}) the fulfillment of conservation laws is obvious, due to independence $T_{22}$ and $T_{33}$ from variables $x^{2}=y$ and $x^{3}=z$.

The system of equations (\ref{E10}), (\ref{E12}), (\ref{HCA1}) and (\ref{HCA2}),
taking into account only the solutions describing the wave process, allows us to find the solutions in the following form
\begin{eqnarray}
\label{WS1S}
&&h_{00}=h_{11}=-\frac{GE^{2}_{0}}{2c^{3}\omega}\left(\frac{L}{2}+x\right)
\sin\left(2\omega\left(t-\frac{x}{c}\right)\right)-
\frac{GE^{2}_{0}}{2c^{3}\omega}\left(\frac{L}{2}-x\right)
\sin\left(2\omega\left(t+\frac{x}{c}\right)\right),\\
\label{WS2S}
&&h_{01}=h_{10}=\frac{GE^{2}_{0}}{2c^{3}\omega}\left(\frac{L}{2}+x\right)
\sin\left(2\omega\left(t-\frac{x}{c}\right)\right)-
\frac{GE^{2}_{0}}{2c^{3}\omega}\left(\frac{L}{2}-x\right)
\sin\left(2\omega\left(t+\frac{x}{c}\right)\right),\\
\label{WS3S}
&&\tilde{h}=h_{22}+h_{33}=\frac{GE^{2}_{0}}{2c^{2}\omega^{2}}\cos\left(2\omega\left(t-\frac{x}{c}\right)\right)-
\frac{GE^{2}_{0}}{2c^{2}\omega^{2}}\cos\left(2\omega\left(t+\frac{x}{c}\right)\right),
\end{eqnarray}
and the system of equations (\ref{E13}) and (\ref{E14}), taking into account (\ref{TEF3S}) and (\ref{WS3S}), allows us to write
\begin{eqnarray}
\label{WS4S}
&&h_{22}=\frac{2GE^{2}_{0}}{c^{2}\omega^{2}}\left[\sin\left(2\omega\left(t-\frac{x}{c}\right)\right)
\sin\left(2\omega\left(t+\frac{x}{c}\right)\right)\right]-
\frac{\alpha GE^{2}_{0}}{2c^{2}\omega^{2}}\left[\cos\left(2\omega\left(t-\frac{x}{c}\right)\right)+
\cos\left(2\omega\left(t+\frac{x}{c}\right)\right)\right],\\
\label{WS5S}
&&h_{33}=-\frac{2GE^{2}_{0}}{c^{2}\omega^{2}}\left[\sin\left(2\omega\left(t-\frac{x}{c}\right)\right)
\sin\left(2\omega\left(t+\frac{x}{c}\right)\right)\right]-
\frac{\beta GE^{2}_{0}}{2c^{2}\omega^{2}}\left[\cos\left(2\omega\left(t-\frac{x}{c}\right)\right)+
\cos\left(2\omega\left(t+\frac{x}{c}\right)\right)\right].
\end{eqnarray}
In obtaining formulas (\ref{WS1S})--(\ref{WS5S}), it was assumed that the gravitational wave is not reflected from the mirrors of the Fabry-Perot resonator.

The density of the energy flow of the gravitational wave, calculated by the formula (\ref{EnFlux}), taking into account the expressions (\ref{WS1S}), (\ref{WS2S}), (\ref{WS4S}) and (\ref{WS5S}) gives
\begin{eqnarray}
\label{EnFlux1S}
\nonumber
&&ct^{01}=\frac{GE^{4}_{0}}{64\pi c\omega^{2}}\Bigg[\left(\frac{L}{2}-x\right)\sin\left(2\omega\left(t-\frac{x}{c}\right)\right)
\cos\left(2\omega\left(t+\frac{x}{c}\right)\right)-\\
&&\left(\frac{L}{2}+x\right)\cos\left(2\omega\left(t-\frac{x}{c}\right)\right)
\sin\left(2\omega\left(t+\frac{x}{c}\right)\right)\Bigg]+ \nonumber
\frac{GE^{4}_{0}}{2\pi c\omega^{2}}\sin\left(2\omega t\right)\sin\left(\frac{2\omega x}{c}\right)+\\
&&\frac{GE^{4}_{0}}{32\pi c\omega^{2}}(\alpha^{2}+\beta^{2})\sin\left(2\omega\left(t-\frac{x}{c}\right)\right)-
\frac{GE^{4}_{0}}{32\pi c\omega^{2}}(\alpha^{2}+\beta^{2})\sin\left(2\omega\left(t+\frac{x}{c}\right)\right).
\end{eqnarray}

If we take into account only the flux density of the gravitational wave (\ref{EnFlux1S}), propagating in the positive direction of the axis $x^{1}=x$, then the energy flux density of the gravitational wave in empty space for $x>L/2$ will be described by formula (\ref{EnFlux1}), and the tensors $h_{22}=-h_{33}$ will be described by expressions (\ref{WS6}) and (\ref{WS7}).

\section{Generating of a gravitational waves at a difference frequency}\label{sect5}

Let the Fabry-Perot resonator simultaneously contain two strong standing electromagnetic waves with close frequencies $\omega_{1}$ and $\omega_{2}$ (for definiteness, we assume that $\omega_{1}>\omega_{2}$), which are described by the formulas
\begin{eqnarray}
\label{EFSS}
&&E_{y}=E_{1}\left[\cos\left(\omega_{1}\left(t-\frac{x}{c}\right)\right)+\cos\left(\omega_{1}\left(t+\frac{x}{c}\right)\right)\right]+
E_{2}\left[\cos\left(\omega_{2}\left(t-\frac{x}{c}\right)\right)+\cos\left(\omega_{2}\left(t+\frac{x}{c}\right)\right)\right],\\
\label{MFSS}
&&H_{z}=H_{1}\left[\cos\left(\omega_{1}\left(t-\frac{x}{c}\right)\right)-\cos\left(\omega_{1}\left(t+\frac{x}{c}\right)\right)\right]+
H_{2}\left[\cos\left(\omega_{2}\left(t-\frac{x}{c}\right)\right)-\cos\left(\omega_{2}\left(t+\frac{x}{c}\right)\right)\right],
\end{eqnarray}
where $E_{1}=H_{1}$ and $E_{2}=H_{2}$.

We write the expressions for the components of the energy-momentum tensor for the terms on the
difference frequency $\tilde{\omega}=\omega_{1}-\omega_{2}$ only as follows
\begin{eqnarray}
\label{TEF1SS}
&&\tilde{T}_{00}=\tilde{T}_{11}=\frac{E_{1}E_{2}}{4\pi}\left[\cos\left(\tilde{\omega}\left(t-\frac{x}{c}\right)\right)+
\cos\left(\tilde{\omega}\left(t+\frac{x}{c}\right)\right)\right],\\
\label{TEF2SS}
&&\tilde{T}_{01}=\tilde{T}_{10}=-\frac{E_{1}E_{2}}{4\pi}\left[\cos\left(\tilde{\omega}\left(t-\frac{x}{c}\right)\right)-
\cos\left(\tilde{\omega}\left(t+\frac{x}{c}\right)\right)\right],
\end{eqnarray}
and the remaining components of this tensor are equal to zero.

Substitution of expressions (\ref{TEF1SS}) and (\ref{TEF2SS}) into the system of equations  (\ref{E10}), (\ref{E12}), (\ref{HCA1}) and (\ref{HCA2}) gives
\begin{eqnarray}
\label{WS1SS}
&&h_{00}=h_{11}=-\frac{4GE_{1}E_{2}}{c^{3}\tilde{\omega}}\left(\frac{L}{2}+x\right)
\sin\left(\tilde{\omega}\left(t-\frac{x}{c}\right)\right)-
\frac{4GE_{1}E_{2}}{c^{3}\tilde{\omega}}\left(\frac{L}{2}-x\right)
\sin\left(\tilde{\omega}\left(t+\frac{x}{c}\right)\right),\\
\label{WS2SS}
&&h_{01}=h_{10}=\frac{4GE_{1}E_{2}}{c^{3}\tilde{\omega}}\left(\frac{L}{2}+x\right)
\sin\left(\tilde{\omega}\left(t-\frac{x}{c}\right)\right)-
\frac{4GE_{1}E_{2}}{c^{3}\tilde{\omega}}\left(\frac{L}{2}-x\right)
\sin\left(\tilde{\omega}\left(t+\frac{x}{c}\right)\right),\\
\label{WS3SS}
&&\tilde{h}=h_{22}+h_{33}=-\frac{4GE_{1}E_{2}}{c^{2}\tilde{\omega}^{2}}
\cos\left(\tilde{\omega}\left(t-\frac{x}{c}\right)\right)-
\frac{4GE_{1}E_{2}}{c^{2}\tilde{\omega}^{2}}\cos\left(\tilde{\omega}\left(t+\frac{x}{c}\right)\right),
\end{eqnarray}

After representing the components $h_{22}$ and $h_{33}$ as
\begin{eqnarray}
\label{WS4SS}
&&h_{22}=-\frac{4\alpha GE_{1}E_{2}}{c^{2}\tilde{\omega}^{2}}
\left[\cos\left(\tilde{\omega}\left(t-\frac{x}{c}\right)\right)+\cos\left(\tilde{\omega}\left(t+\frac{x}{c}\right)\right)\right],\\
\label{WS5SS}
&&h_{33}=-\frac{4\beta GE_{1}E_{2}}{c^{2}\tilde{\omega}^{2}}
\left[\cos\left(\tilde{\omega}\left(t-\frac{x}{c}\right)\right)+\cos\left(\tilde{\omega}\left(t+\frac{x}{c}\right)\right)\right].
\end{eqnarray}
and performing transformations similar to (\ref{EnFlux1}) and (\ref{WS6}), we find the components $h_{22}=-h_{33}$ in empty space for $x>L/2$ in the following form
\begin{equation}
\label{WS6SS}
h_{22}=-h_{33}=\pm\frac{4GE_{1}E_{2}\sqrt{\alpha^{2}+\beta^{2}}}{\sqrt{2}c^{2}\tilde{\omega}^{2}}
\cos\left(\tilde{\omega}\left(t-\frac{x}{c}\right)\right),
\end{equation}
and for $\alpha=\beta=1/2$ we finally obtain
\begin{equation}
\label{WS7SS}
h_{22}=-h_{33}=\pm\frac{2GE_{1}E_{2}}{c^{2}\tilde{\omega}^{2}}
\cos\left(\tilde{\omega}\left(t-\frac{x}{c}\right)\right).
\end{equation}

A fundamental feature of the obtained solution (\ref{WS6SS}) from (\ref{WS6}) is the possibility of generating the low-frequency gravitational waves with $\tilde{\omega}\ll\omega_{1,2}$ by means of a high-frequency standing electromagnetic waves.

\section{Conclusion}
In this paper, we investigated a bound states of electromagnetic and gravitational waves based on Einstein-Maxwell equations.
The gravitational-wave part of these states included both the longitudinal and transverse components, which were transformed into transverse gravitational waves in vacuum after absorption of an electromagnetic wave.

Thus, the coupled gravitational waves exist in empty space when electromagnetic waves propagate in it. This, in particular, leads to the fact that, in addition to transverse gravitational waves arising from merging or due to rotation of the astrophysical objects, such a gravitational waves will be generated and propagated along with electromagnetic radiation accompanying these processes.

The subsequent analysis of a such bound states confirms the possibility of the existence of longitudinal-transverse gravitational waves coupled with a strong propagating and standing electromagnetic waves as well, which form free transverse gravitational waves in empty space.

It was also received, that two standing electromagnetic waves in the Fabry-Perot resonator having close frequencies lead to the emission of transverse gravitational waves at the difference frequency. The analysis of the possibility of generating and detecting coupled gravitational waves allows us to move on to the creation of experimental facilities and to conduct preliminary measurements on them.

\begin{acknowledgments}
The study was funded by a grant from the Russian Science Foundation (project No. 19-12-00242).
\end{acknowledgments}

\bibliography{ref}

\begin{thebibliography}{42}%
\makeatletter
\providecommand \@ifxundefined [1]{%
 \@ifx{#1\undefined}
}%
\providecommand \@ifnum [1]{%
 \ifnum #1\expandafter \@firstoftwo
 \else \expandafter \@secondoftwo
 \fi
}%
\providecommand \@ifx [1]{%
 \ifx #1\expandafter \@firstoftwo
 \else \expandafter \@secondoftwo
 \fi
}%
\providecommand \natexlab [1]{#1}%
\providecommand \enquote  [1]{``#1''}%
\providecommand \bibnamefont  [1]{#1}%
\providecommand \bibfnamefont [1]{#1}%
\providecommand \citenamefont [1]{#1}%
\providecommand \href@noop [0]{\@secondoftwo}%
\providecommand \href [0]{\begingroup \@sanitize@url \@href}%
\providecommand \@href[1]{\@@startlink{#1}\@@href}%
\providecommand \@@href[1]{\endgroup#1\@@endlink}%
\providecommand \@sanitize@url [0]{\catcode `\\12\catcode `\$12\catcode
  `\&12\catcode `\#12\catcode `\^12\catcode `\_12\catcode `\%12\relax}%
\providecommand \@@startlink[1]{}%
\providecommand \@@endlink[0]{}%
\providecommand \url  [0]{\begingroup\@sanitize@url \@url }%
\providecommand \@url [1]{\endgroup\@href {#1}{\urlprefix }}%
\providecommand \urlprefix  [0]{URL }%
\providecommand \Eprint [0]{\href }%
\providecommand \doibase [0]{http://dx.doi.org/}%
\providecommand \selectlanguage [0]{\@gobble}%
\providecommand \bibinfo  [0]{\@secondoftwo}%
\providecommand \bibfield  [0]{\@secondoftwo}%
\providecommand \translation [1]{[#1]}%
\providecommand \BibitemOpen [0]{}%
\providecommand \bibitemStop [0]{}%
\providecommand \bibitemNoStop [0]{.\EOS\space}%
\providecommand \EOS [0]{\spacefactor3000\relax}%
\providecommand \BibitemShut  [1]{\csname bibitem#1\endcsname}%
\let\auto@bib@innerbib\@empty
%</preamble>
\bibitem [{\citenamefont {Einstein}(1987)}]{Einstein}%
  \BibitemOpen
  \bibfield  {author} {\bibinfo {author} {\bibfnamefont {A.}~\bibnamefont
  {Einstein}},\ }\href {http://einsteinpapers.press.princeton.edu} {\emph
  {\bibinfo {title} {{The Collected Papers of Albert Einstein (CPAE)}}}}\
  (\bibinfo  {publisher} {Princeton University Press},\ \bibinfo {address}
  {Princeton},\ \bibinfo {year} {1987})\BibitemShut {NoStop}%
\bibitem [{\citenamefont {Eddington}(1923)}]{Eddington}%
  \BibitemOpen
  \bibfield  {author} {\bibinfo {author} {\bibfnamefont {A.~S.}\ \bibnamefont
  {Eddington}},\ }\href@noop {} {\emph {\bibinfo {title} {{The mathematical
  theory of relativity}}}}\ (\bibinfo  {publisher} {Cambridge University
  Press},\ \bibinfo {address} {Cambridge},\ \bibinfo {year} {1923})\BibitemShut
  {NoStop}%
\bibitem [{\citenamefont {Landau}\ and\ \citenamefont
  {Lifshitz}(1987)}]{Landau}%
  \BibitemOpen
  \bibfield  {author} {\bibinfo {author} {\bibfnamefont {L.~D.}\ \bibnamefont
  {Landau}}\ and\ \bibinfo {author} {\bibfnamefont {E.~M.}\ \bibnamefont
  {Lifshitz}},\ }\href
  {https://www.elsevier.com/books/the-classical-theory-of-fields/landau/978-0-08-050349-3}
  {\emph {\bibinfo {title} {{The Classical Theory of Fields}}}}\ (\bibinfo
  {publisher} {Butterworth-Heinemann},\ \bibinfo {year} {1987})\BibitemShut
  {NoStop}%
\bibitem [{\citenamefont {Weber}(1961)}]{Weber}%
  \BibitemOpen
  \bibfield  {author} {\bibinfo {author} {\bibfnamefont {J.}~\bibnamefont
  {Weber}},\ }\href@noop {} {\emph {\bibinfo {title} {{General relativity and
  gravitational waves}}}}\ (\bibinfo  {publisher} {Interscience},\ \bibinfo
  {address} {New York},\ \bibinfo {year} {1961})\BibitemShut {NoStop}%
\bibitem [{\citenamefont {Maggiore}(2008)}]{Maggiore:2018sht}%
  \BibitemOpen
  \bibfield  {author} {\bibinfo {author} {\bibfnamefont {M.}~\bibnamefont
  {Maggiore}},\ }\href@noop {} {\emph {\bibinfo {title} {{Gravitational Waves:
  Volume 1: Theory and Experiments}}}}\ (\bibinfo  {publisher} {Oxford
  University Press},\ \bibinfo {address} {Oxford},\ \bibinfo {year}
  {2008})\BibitemShut {NoStop}%
\bibitem [{\citenamefont {Abbott}\ \emph
  {et~al.}(2018{\natexlab{a}})\citenamefont {Abbott} \emph
  {et~al.}}]{Aasi:2013wya}%
  \BibitemOpen
  \bibfield  {author} {\bibinfo {author} {\bibfnamefont {B.~P.}\ \bibnamefont
  {Abbott}} \emph {et~al.} (\bibinfo {collaboration} {KAGRA, LIGO Scientific,
  VIRGO}),\ }\href {\doibase 10.1007/s41114-018-0012-9, 10.1007/lrr-2016-1}
  {\bibfield  {journal} {\bibinfo  {journal} {Living Rev. Rel.}\ }\textbf
  {\bibinfo {volume} {21}},\ \bibinfo {pages} {3} (\bibinfo {year}
  {2018}{\natexlab{a}})},\ \Eprint {http://arxiv.org/abs/1304.0670}
  {arXiv:1304.0670 [gr-qc]} \BibitemShut {NoStop}%
%%CITATION = ARXIV:1304.0670;%%
\bibitem [{\citenamefont {Abbott}\ \emph
  {et~al.}(2016{\natexlab{a}})\citenamefont {Abbott} \emph
  {et~al.}}]{Abbott:2016blz}%
  \BibitemOpen
  \bibfield  {author} {\bibinfo {author} {\bibfnamefont {B.~P.}\ \bibnamefont
  {Abbott}} \emph {et~al.} (\bibinfo {collaboration} {LIGO Scientific,
  Virgo}),\ }\href {\doibase 10.1103/PhysRevLett.116.061102} {\bibfield
  {journal} {\bibinfo  {journal} {Phys. Rev. Lett.}\ }\textbf {\bibinfo
  {volume} {116}},\ \bibinfo {pages} {061102} (\bibinfo {year}
  {2016}{\natexlab{a}})},\ \Eprint {http://arxiv.org/abs/1602.03837}
  {arXiv:1602.03837 [gr-qc]} \BibitemShut {NoStop}%
%%CITATION = ARXIV:1602.03837;%%
\bibitem [{\citenamefont {Abbott}\ \emph
  {et~al.}(2016{\natexlab{b}})\citenamefont {Abbott} \emph
  {et~al.}}]{Abbott:2016nmj}%
  \BibitemOpen
  \bibfield  {author} {\bibinfo {author} {\bibfnamefont {B.~P.}\ \bibnamefont
  {Abbott}} \emph {et~al.} (\bibinfo {collaboration} {LIGO Scientific,
  Virgo}),\ }\href {\doibase 10.1103/PhysRevLett.116.241103} {\bibfield
  {journal} {\bibinfo  {journal} {Phys. Rev. Lett.}\ }\textbf {\bibinfo
  {volume} {116}},\ \bibinfo {pages} {241103} (\bibinfo {year}
  {2016}{\natexlab{b}})},\ \Eprint {http://arxiv.org/abs/1606.04855}
  {arXiv:1606.04855 [gr-qc]} \BibitemShut {NoStop}%
%%CITATION = ARXIV:1606.04855;%%
\bibitem [{\citenamefont {Abbott}\ \emph {et~al.}(2017)\citenamefont {Abbott}
  \emph {et~al.}}]{Monitor:2017mdv}%
  \BibitemOpen
  \bibfield  {author} {\bibinfo {author} {\bibfnamefont {B.~P.}\ \bibnamefont
  {Abbott}} \emph {et~al.} (\bibinfo {collaboration} {LIGO Scientific, Virgo,
  Fermi-GBM, INTEGRAL}),\ }\href {\doibase 10.3847/2041-8213/aa920c} {\bibfield
   {journal} {\bibinfo  {journal} {Astrophys. J.}\ }\textbf {\bibinfo {volume}
  {848}},\ \bibinfo {pages} {L13} (\bibinfo {year} {2017})},\ \Eprint
  {http://arxiv.org/abs/1710.05834} {arXiv:1710.05834 [astro-ph.HE]}
  \BibitemShut {NoStop}%
%%CITATION = ARXIV:1710.05834;%%
\bibitem [{\citenamefont {Gertsenshtein}\ and\ \citenamefont
  {Pustovoit}(1962)}]{Gertsenshtein:1962kfm}%
  \BibitemOpen
  \bibfield  {author} {\bibinfo {author} {\bibfnamefont {M.~E.}\ \bibnamefont
  {Gertsenshtein}}\ and\ \bibinfo {author} {\bibfnamefont {V.~I.}\ \bibnamefont
  {Pustovoit}},\ }\href@noop {} {\bibfield  {journal} {\bibinfo  {journal}
  {JETP}\ }\textbf {\bibinfo {volume} {43}},\ \bibinfo {pages} {605} (\bibinfo
  {year} {1962})},\ \bibinfo {note} {[Translate M. E. Gertsenshtein and V. I.
  Pustovoit, Sov. Phys. JETP, 16, 433 (1963)]}\BibitemShut {NoStop}%
\bibitem [{\citenamefont {Liang}\ \emph {et~al.}(2017)\citenamefont {Liang},
  \citenamefont {Gong}, \citenamefont {Hou},\ and\ \citenamefont
  {Liu}}]{Liang:2017ahj}%
  \BibitemOpen
  \bibfield  {author} {\bibinfo {author} {\bibfnamefont {D.}~\bibnamefont
  {Liang}}, \bibinfo {author} {\bibfnamefont {Y.}~\bibnamefont {Gong}},
  \bibinfo {author} {\bibfnamefont {S.}~\bibnamefont {Hou}}, \ and\ \bibinfo
  {author} {\bibfnamefont {Y.}~\bibnamefont {Liu}},\ }\href {\doibase
  10.1103/PhysRevD.95.104034} {\bibfield  {journal} {\bibinfo  {journal} {Phys.
  Rev.}\ }\textbf {\bibinfo {volume} {D95}},\ \bibinfo {pages} {104034}
  (\bibinfo {year} {2017})},\ \Eprint {http://arxiv.org/abs/1701.05998}
  {arXiv:1701.05998 [gr-qc]} \BibitemShut {NoStop}%
%%CITATION = ARXIV:1701.05998;%%
\bibitem [{\citenamefont {Hou}\ \emph {et~al.}(2018)\citenamefont {Hou},
  \citenamefont {Gong},\ and\ \citenamefont {Liu}}]{Hou:2017bqj}%
  \BibitemOpen
  \bibfield  {author} {\bibinfo {author} {\bibfnamefont {S.}~\bibnamefont
  {Hou}}, \bibinfo {author} {\bibfnamefont {Y.}~\bibnamefont {Gong}}, \ and\
  \bibinfo {author} {\bibfnamefont {Y.}~\bibnamefont {Liu}},\ }\href {\doibase
  10.1140/epjc/s10052-018-5869-y} {\bibfield  {journal} {\bibinfo  {journal}
  {Eur. Phys. J.}\ }\textbf {\bibinfo {volume} {C78}},\ \bibinfo {pages} {378}
  (\bibinfo {year} {2018})},\ \Eprint {http://arxiv.org/abs/1704.01899}
  {arXiv:1704.01899 [gr-qc]} \BibitemShut {NoStop}%
%%CITATION = ARXIV:1704.01899;%%
\bibitem [{\citenamefont {Hyun}\ \emph {et~al.}(2019)\citenamefont {Hyun},
  \citenamefont {Kim},\ and\ \citenamefont {Lee}}]{Hyun:2018pgn}%
  \BibitemOpen
  \bibfield  {author} {\bibinfo {author} {\bibfnamefont {Y.-H.}\ \bibnamefont
  {Hyun}}, \bibinfo {author} {\bibfnamefont {Y.}~\bibnamefont {Kim}}, \ and\
  \bibinfo {author} {\bibfnamefont {S.}~\bibnamefont {Lee}},\ }\href {\doibase
  10.1103/PhysRevD.99.124002} {\bibfield  {journal} {\bibinfo  {journal} {Phys.
  Rev.}\ }\textbf {\bibinfo {volume} {D99}},\ \bibinfo {pages} {124002}
  (\bibinfo {year} {2019})},\ \Eprint {http://arxiv.org/abs/1810.09316}
  {arXiv:1810.09316 [gr-qc]} \BibitemShut {NoStop}%
%%CITATION = ARXIV:1810.09316;%%
\bibitem [{\citenamefont {Wagle}\ \emph {et~al.}(2019)\citenamefont {Wagle},
  \citenamefont {Saffer},\ and\ \citenamefont {Yunes}}]{Wagle:2019mdq}%
  \BibitemOpen
  \bibfield  {author} {\bibinfo {author} {\bibfnamefont {P.}~\bibnamefont
  {Wagle}}, \bibinfo {author} {\bibfnamefont {A.}~\bibnamefont {Saffer}}, \
  and\ \bibinfo {author} {\bibfnamefont {N.}~\bibnamefont {Yunes}},\ }\href
  {\doibase 10.1103/PhysRevD.100.124007} {\  (\bibinfo {year} {2019}),\
  10.1103/PhysRevD.100.124007},\ \bibinfo {note} {[Phys.
  Rev.D100,no.12,124007(2019)]},\ \Eprint {http://arxiv.org/abs/1910.04800}
  {arXiv:1910.04800 [gr-qc]} \BibitemShut {NoStop}%
%%CITATION = ARXIV:1910.04800;%%
\bibitem [{\citenamefont
  {Shankaranarayanan}(2019)}]{Shankaranarayanan:2019yjx}%
  \BibitemOpen
  \bibfield  {author} {\bibinfo {author} {\bibfnamefont {S.}~\bibnamefont
  {Shankaranarayanan}},\ }\href {\doibase 10.1142/S0218271819440206} {\bibfield
   {journal} {\bibinfo  {journal} {Int. J. Mod. Phys.}\ }\textbf {\bibinfo
  {volume} {D28}},\ \bibinfo {pages} {1944020} (\bibinfo {year} {2019})},\
  \Eprint {http://arxiv.org/abs/1905.03943} {arXiv:1905.03943 [gr-qc]}
  \BibitemShut {NoStop}%
%%CITATION = ARXIV:1905.03943;%%
\bibitem [{\citenamefont {Nishizawa}\ \emph {et~al.}(2009)\citenamefont
  {Nishizawa}, \citenamefont {Taruya}, \citenamefont {Hayama}, \citenamefont
  {Kawamura},\ and\ \citenamefont {Sakagami}}]{Nishizawa:2009bf}%
  \BibitemOpen
  \bibfield  {author} {\bibinfo {author} {\bibfnamefont {A.}~\bibnamefont
  {Nishizawa}}, \bibinfo {author} {\bibfnamefont {A.}~\bibnamefont {Taruya}},
  \bibinfo {author} {\bibfnamefont {K.}~\bibnamefont {Hayama}}, \bibinfo
  {author} {\bibfnamefont {S.}~\bibnamefont {Kawamura}}, \ and\ \bibinfo
  {author} {\bibfnamefont {M.}~\bibnamefont {Sakagami}},\ }\href {\doibase
  10.1103/PhysRevD.79.082002} {\bibfield  {journal} {\bibinfo  {journal} {Phys.
  Rev.}\ }\textbf {\bibinfo {volume} {D79}},\ \bibinfo {pages} {082002}
  (\bibinfo {year} {2009})},\ \Eprint {http://arxiv.org/abs/0903.0528}
  {arXiv:0903.0528 [astro-ph.CO]} \BibitemShut {NoStop}%
%%CITATION = ARXIV:0903.0528;%%
\bibitem [{\citenamefont {Philippoz}\ and\ \citenamefont
  {Jetzer}(2017)}]{Philippoz:2017ywb}%
  \BibitemOpen
  \bibfield  {author} {\bibinfo {author} {\bibfnamefont {L.}~\bibnamefont
  {Philippoz}}\ and\ \bibinfo {author} {\bibfnamefont {P.}~\bibnamefont
  {Jetzer}},\ }\href {\doibase 10.1088/1742-6596/840/1/012057} {\bibfield
  {journal} {\bibinfo  {journal} {J. Phys. Conf. Ser.}\ }\textbf {\bibinfo
  {volume} {840}},\ \bibinfo {pages} {012057} (\bibinfo {year}
  {2017})}\BibitemShut {NoStop}%
%%CITATION = 00462,840,012057;%%
\bibitem [{\citenamefont {Abbott}\ \emph
  {et~al.}(2018{\natexlab{b}})\citenamefont {Abbott} \emph
  {et~al.}}]{Abbott:2018utx}%
  \BibitemOpen
  \bibfield  {author} {\bibinfo {author} {\bibfnamefont {B.~P.}\ \bibnamefont
  {Abbott}} \emph {et~al.} (\bibinfo {collaboration} {LIGO Scientific,
  Virgo}),\ }\href {\doibase 10.1103/PhysRevLett.120.201102} {\bibfield
  {journal} {\bibinfo  {journal} {Phys. Rev. Lett.}\ }\textbf {\bibinfo
  {volume} {120}},\ \bibinfo {pages} {201102} (\bibinfo {year}
  {2018}{\natexlab{b}})},\ \Eprint {http://arxiv.org/abs/1802.10194}
  {arXiv:1802.10194 [gr-qc]} \BibitemShut {NoStop}%
%%CITATION = ARXIV:1802.10194;%%
\bibitem [{\citenamefont {Takeda}\ \emph {et~al.}(2018)\citenamefont {Takeda},
  \citenamefont {Nishizawa}, \citenamefont {Michimura}, \citenamefont {Nagano},
  \citenamefont {Komori}, \citenamefont {Ando},\ and\ \citenamefont
  {Hayama}}]{Takeda:2018uai}%
  \BibitemOpen
  \bibfield  {author} {\bibinfo {author} {\bibfnamefont {H.}~\bibnamefont
  {Takeda}}, \bibinfo {author} {\bibfnamefont {A.}~\bibnamefont {Nishizawa}},
  \bibinfo {author} {\bibfnamefont {Y.}~\bibnamefont {Michimura}}, \bibinfo
  {author} {\bibfnamefont {K.}~\bibnamefont {Nagano}}, \bibinfo {author}
  {\bibfnamefont {K.}~\bibnamefont {Komori}}, \bibinfo {author} {\bibfnamefont
  {M.}~\bibnamefont {Ando}}, \ and\ \bibinfo {author} {\bibfnamefont
  {K.}~\bibnamefont {Hayama}},\ }\href {\doibase 10.1103/PhysRevD.98.022008}
  {\bibfield  {journal} {\bibinfo  {journal} {Phys. Rev.}\ }\textbf {\bibinfo
  {volume} {D98}},\ \bibinfo {pages} {022008} (\bibinfo {year} {2018})},\
  \Eprint {http://arxiv.org/abs/1806.02182} {arXiv:1806.02182 [gr-qc]}
  \BibitemShut {NoStop}%
%%CITATION = ARXIV:1806.02182;%%
\bibitem [{\citenamefont {Grishchuk}\ and\ \citenamefont
  {Sazhin}(1973)}]{Grishchuk:1973qz}%
  \BibitemOpen
  \bibfield  {author} {\bibinfo {author} {\bibfnamefont {L.~P.}\ \bibnamefont
  {Grishchuk}}\ and\ \bibinfo {author} {\bibfnamefont {M.~V.}\ \bibnamefont
  {Sazhin}},\ }\href@noop {} {\bibfield  {journal} {\bibinfo  {journal} {Zh.
  Eksp. Teor. Fiz.}\ }\textbf {\bibinfo {volume} {65}},\ \bibinfo {pages} {441}
  (\bibinfo {year} {1973})}\BibitemShut {NoStop}%
%%CITATION = ZETFA,65,441;%%
\bibitem [{\citenamefont {Denisov}(1977)}]{Denisov}%
  \BibitemOpen
  \bibfield  {author} {\bibinfo {author} {\bibfnamefont {V.~I.}\ \bibnamefont
  {Denisov}},\ }\href@noop {} {\bibfield  {journal} {\bibinfo  {journal}
  {Moscow University Physics Bulletin}\ }\textbf {\bibinfo {volume} {32}},\
  \bibinfo {pages} {41} (\bibinfo {year} {1977})}\BibitemShut {NoStop}%
\bibitem [{\citenamefont {Nikishov}\ and\ \citenamefont
  {Ritus}(2011)}]{Nikishov:2010zz}%
  \BibitemOpen
  \bibfield  {author} {\bibinfo {author} {\bibfnamefont {A.~I.}\ \bibnamefont
  {Nikishov}}\ and\ \bibinfo {author} {\bibfnamefont {V.~I.}\ \bibnamefont
  {Ritus}},\ }\href {\doibase 10.3367/UFNe.0180.201011b.1135} {\bibfield
  {journal} {\bibinfo  {journal} {Phys. Usp.}\ }\textbf {\bibinfo {volume}
  {53}},\ \bibinfo {pages} {1093} (\bibinfo {year} {2011})},\ \bibinfo {note}
  {[Usp. Fiz. Nauk180,1135(2010)]}\BibitemShut {NoStop}%
%%CITATION = PHUSE,53,1093;%%
\bibitem [{\citenamefont {Hough}\ \emph {et~al.}(2005)\citenamefont {Hough},
  \citenamefont {Rowan},\ and\ \citenamefont {Sathyaprakash}}]{Hough:2005pf}%
  \BibitemOpen
  \bibfield  {author} {\bibinfo {author} {\bibfnamefont {J.}~\bibnamefont
  {Hough}}, \bibinfo {author} {\bibfnamefont {S.}~\bibnamefont {Rowan}}, \ and\
  \bibinfo {author} {\bibfnamefont {B.~S.}\ \bibnamefont {Sathyaprakash}},\
  }\href {\doibase 10.1088/0953-4075/38/9/004} {\bibfield  {journal} {\bibinfo
  {journal} {J. Phys.}\ }\textbf {\bibinfo {volume} {B38}},\ \bibinfo {pages}
  {S497} (\bibinfo {year} {2005})},\ \Eprint
  {http://arxiv.org/abs/gr-qc/0501007} {arXiv:gr-qc/0501007 [gr-qc]}
  \BibitemShut {NoStop}%
%%CITATION = GR-QC/0501007;%%
\bibitem [{\citenamefont {Chen}\ \emph {et~al.}(2017)\citenamefont {Chen},
  \citenamefont {Nester},\ and\ \citenamefont {Ni}}]{Chen:2016isk}%
  \BibitemOpen
  \bibfield  {author} {\bibinfo {author} {\bibfnamefont {C.-M.}\ \bibnamefont
  {Chen}}, \bibinfo {author} {\bibfnamefont {J.~M.}\ \bibnamefont {Nester}}, \
  and\ \bibinfo {author} {\bibfnamefont {W.-T.}\ \bibnamefont {Ni}},\ }\href
  {\doibase 10.1016/j.cjph.2016.10.014} {\bibfield  {journal} {\bibinfo
  {journal} {Chin. J. Phys.}\ }\textbf {\bibinfo {volume} {55}},\ \bibinfo
  {pages} {142} (\bibinfo {year} {2017})},\ \Eprint
  {http://arxiv.org/abs/1610.08803} {arXiv:1610.08803 [gr-qc]} \BibitemShut
  {NoStop}%
%%CITATION = ARXIV:1610.08803;%%
\bibitem [{\citenamefont {Pustovoit}(2016)}]{Pustovoit:2016zyt}%
  \BibitemOpen
  \bibfield  {author} {\bibinfo {author} {\bibfnamefont {V.~I.}\ \bibnamefont
  {Pustovoit}},\ }\href {\doibase 10.3367/UFNe.2016.03.037900} {\bibfield
  {journal} {\bibinfo  {journal} {Phys. Usp.}\ }\textbf {\bibinfo {volume}
  {59}},\ \bibinfo {pages} {1034} (\bibinfo {year} {2016})},\ \bibinfo {note}
  {[Usp. Fiz. Nauk186,no.10,1133(2016)]}\BibitemShut {NoStop}%
%%CITATION = PHUSE,59,1034;%%
\bibitem [{\citenamefont {Rudenko}(2017)}]{Rudenko:2017asr}%
  \BibitemOpen
  \bibfield  {author} {\bibinfo {author} {\bibfnamefont {V.~N.}\ \bibnamefont
  {Rudenko}},\ }\href {\doibase 10.3367/UFNe.2016.11.038088} {\bibfield
  {journal} {\bibinfo  {journal} {Phys. Usp.}\ }\textbf {\bibinfo {volume}
  {60}},\ \bibinfo {pages} {830} (\bibinfo {year} {2017})}\BibitemShut
  {NoStop}%
%%CITATION = PHUSE,60,830;%%
\bibitem [{\citenamefont {Grishchuk}(1977)}]{Grishchuk_1977}%
  \BibitemOpen
  \bibfield  {author} {\bibinfo {author} {\bibfnamefont {L.~P.}\ \bibnamefont
  {Grishchuk}},\ }\href {\doibase 10.1070/pu1977v020n04abeh005327} {\bibfield
  {journal} {\bibinfo  {journal} {Soviet Physics Uspekhi}\ }\textbf {\bibinfo
  {volume} {20}},\ \bibinfo {pages} {319} (\bibinfo {year} {1977})}\BibitemShut
  {NoStop}%
\bibitem [{\citenamefont {Eddington}(1922)}]{Eddington1}%
  \BibitemOpen
  \bibfield  {author} {\bibinfo {author} {\bibfnamefont {A.~S.}\ \bibnamefont
  {Eddington}},\ }\href@noop {} {\bibfield  {journal} {\bibinfo  {journal}
  {Proc. Roy. Soc. London A}\ }\textbf {\bibinfo {volume} {102}},\ \bibinfo
  {pages} {268} (\bibinfo {year} {1922})}\BibitemShut {NoStop}%
\bibitem [{\citenamefont {Gertsenshtein}(1962)}]{Gertsenshtein1962}%
  \BibitemOpen
  \bibfield  {author} {\bibinfo {author} {\bibfnamefont {M.~E.}\ \bibnamefont
  {Gertsenshtein}},\ }\href@noop {} {\bibfield  {journal} {\bibinfo  {journal}
  {Sov. Phys. JETP}\ }\textbf {\bibinfo {volume} {14}},\ \bibinfo {pages} {84}
  (\bibinfo {year} {1962})}\BibitemShut {NoStop}%
\bibitem [{\citenamefont {Zeldovich}(1974)}]{Zel'dovich1974}%
  \BibitemOpen
  \bibfield  {author} {\bibinfo {author} {\bibfnamefont {Y.~B.}\ \bibnamefont
  {Zeldovich}},\ }\href@noop {} {\bibfield  {journal} {\bibinfo  {journal}
  {Sov. Phys. JETP}\ }\textbf {\bibinfo {volume} {38}},\ \bibinfo {pages} {652}
  (\bibinfo {year} {1974})}\BibitemShut {NoStop}%
\bibitem [{\citenamefont {Gerlach}(1974)}]{Gerlach:1974zz}%
  \BibitemOpen
  \bibfield  {author} {\bibinfo {author} {\bibfnamefont {U.~H.}\ \bibnamefont
  {Gerlach}},\ }\href {\doibase 10.1103/PhysRevLett.32.1023} {\bibfield
  {journal} {\bibinfo  {journal} {Phys. Rev. Lett.}\ }\textbf {\bibinfo
  {volume} {32}},\ \bibinfo {pages} {1023} (\bibinfo {year}
  {1974})}\BibitemShut {NoStop}%
%%CITATION = PRLTA,32,1023;%%
\bibitem [{\citenamefont {Zeldovich}\ and\ \citenamefont
  {Novikov}(1983)}]{Zeldovich1983}%
  \BibitemOpen
  \bibfield  {author} {\bibinfo {author} {\bibfnamefont {Y.~B.}\ \bibnamefont
  {Zeldovich}}\ and\ \bibinfo {author} {\bibfnamefont {I.~D.}\ \bibnamefont
  {Novikov}},\ }\href@noop {} {\emph {\bibinfo {title} {{Relativistic
  Astrophysics Vol. 2}}}}\ (\bibinfo  {publisher} {University of Chicago
  Press},\ \bibinfo {address} {Chicago},\ \bibinfo {year} {1983})\BibitemShut
  {NoStop}%
\bibitem [{\citenamefont {Raffelt}\ and\ \citenamefont
  {Stodolsky}(1988)}]{Raffelt:1987im}%
  \BibitemOpen
  \bibfield  {author} {\bibinfo {author} {\bibfnamefont {G.}~\bibnamefont
  {Raffelt}}\ and\ \bibinfo {author} {\bibfnamefont {L.}~\bibnamefont
  {Stodolsky}},\ }\href {\doibase 10.1103/PhysRevD.37.1237} {\bibfield
  {journal} {\bibinfo  {journal} {Phys. Rev.}\ }\textbf {\bibinfo {volume}
  {D37}},\ \bibinfo {pages} {1237} (\bibinfo {year} {1988})}\BibitemShut
  {NoStop}%
%%CITATION = PHRVA,D37,1237;%%
\bibitem [{\citenamefont {Pustovoit}\ and\ \citenamefont
  {Chernozatonsky}(1981)}]{Pustovoit:1981za}%
  \BibitemOpen
  \bibfield  {author} {\bibinfo {author} {\bibfnamefont {V.~I.}\ \bibnamefont
  {Pustovoit}}\ and\ \bibinfo {author} {\bibfnamefont {L.~A.}\ \bibnamefont
  {Chernozatonsky}},\ }\href@noop {} {\bibfield  {journal} {\bibinfo  {journal}
  {Zh. Eksp. Teor. Fiz.}\ }\textbf {\bibinfo {volume} {34}},\ \bibinfo {pages}
  {241} (\bibinfo {year} {1981})}\BibitemShut {NoStop}%
%%CITATION = ZETFA,34,241;%%
\bibitem [{\citenamefont {Fargion}(1995)}]{Fargion:1995mm}%
  \BibitemOpen
  \bibfield  {author} {\bibinfo {author} {\bibfnamefont {D.}~\bibnamefont
  {Fargion}},\ }\href@noop {} {\bibfield  {journal} {\bibinfo  {journal} {Grav.
  Cosmol.}\ }\textbf {\bibinfo {volume} {1}},\ \bibinfo {pages} {301} (\bibinfo
  {year} {1995})},\ \Eprint {http://arxiv.org/abs/astro-ph/9604047}
  {arXiv:astro-ph/9604047 [astro-ph]} \BibitemShut {NoStop}%
%%CITATION = ASTRO-PH/9604047;%%
\bibitem [{\citenamefont {Marklund}\ \emph {et~al.}(2000)\citenamefont
  {Marklund}, \citenamefont {Brodin},\ and\ \citenamefont
  {Dunsby}}]{Marklund:1999sp}%
  \BibitemOpen
  \bibfield  {author} {\bibinfo {author} {\bibfnamefont {M.}~\bibnamefont
  {Marklund}}, \bibinfo {author} {\bibfnamefont {G.}~\bibnamefont {Brodin}}, \
  and\ \bibinfo {author} {\bibfnamefont {P.~K.~S.}\ \bibnamefont {Dunsby}},\
  }\href {\doibase 10.1086/308957} {\bibfield  {journal} {\bibinfo  {journal}
  {Astrophys. J.}\ }\textbf {\bibinfo {volume} {536}},\ \bibinfo {pages} {875}
  (\bibinfo {year} {2000})},\ \Eprint {http://arxiv.org/abs/astro-ph/9907350}
  {arXiv:astro-ph/9907350 [astro-ph]} \BibitemShut {NoStop}%
%%CITATION = ASTRO-PH/9907350;%%
\bibitem [{\citenamefont {Dolgov}\ and\ \citenamefont
  {Ejlli}(2012)}]{Dolgov:2012be}%
  \BibitemOpen
  \bibfield  {author} {\bibinfo {author} {\bibfnamefont {A.~D.}\ \bibnamefont
  {Dolgov}}\ and\ \bibinfo {author} {\bibfnamefont {D.}~\bibnamefont {Ejlli}},\
  }\href {\doibase 10.1088/1475-7516/2012/12/003} {\bibfield  {journal}
  {\bibinfo  {journal} {JCAP}\ }\textbf {\bibinfo {volume} {1212}},\ \bibinfo
  {pages} {003} (\bibinfo {year} {2012})},\ \Eprint
  {http://arxiv.org/abs/1211.0500} {arXiv:1211.0500 [gr-qc]} \BibitemShut
  {NoStop}%
%%CITATION = ARXIV:1211.0500;%%
\bibitem [{\citenamefont {Kolosnitsyn}\ and\ \citenamefont
  {Rudenko}(2015)}]{Kolosnitsyn:2015zua}%
  \BibitemOpen
  \bibfield  {author} {\bibinfo {author} {\bibfnamefont {N.~I.}\ \bibnamefont
  {Kolosnitsyn}}\ and\ \bibinfo {author} {\bibfnamefont {V.~N.}\ \bibnamefont
  {Rudenko}},\ }\href {\doibase 10.1088/0031-8949/90/7/074059} {\bibfield
  {journal} {\bibinfo  {journal} {Phys. Scripta}\ }\textbf {\bibinfo {volume}
  {90}},\ \bibinfo {pages} {074059} (\bibinfo {year} {2015})},\ \Eprint
  {http://arxiv.org/abs/1504.06548} {arXiv:1504.06548 [gr-qc]} \BibitemShut
  {NoStop}%
%%CITATION = ARXIV:1504.06548;%%
\bibitem [{\citenamefont {Dolgov}\ and\ \citenamefont
  {Postnov}(2017)}]{Dolgov:2017bpj}%
  \BibitemOpen
  \bibfield  {author} {\bibinfo {author} {\bibfnamefont {A.}~\bibnamefont
  {Dolgov}}\ and\ \bibinfo {author} {\bibfnamefont {K.}~\bibnamefont
  {Postnov}},\ }\href {\doibase 10.1088/1475-7516/2017/09/018} {\bibfield
  {journal} {\bibinfo  {journal} {JCAP}\ }\textbf {\bibinfo {volume} {1709}},\
  \bibinfo {pages} {018} (\bibinfo {year} {2017})},\ \Eprint
  {http://arxiv.org/abs/1706.05519} {arXiv:1706.05519 [astro-ph.HE]}
  \BibitemShut {NoStop}%
%%CITATION = ARXIV:1706.05519;%%
\bibitem [{\citenamefont {Ejlli}\ and\ \citenamefont
  {Thandlam}(2019)}]{Ejlli:2018hke}%
  \BibitemOpen
  \bibfield  {author} {\bibinfo {author} {\bibfnamefont {D.}~\bibnamefont
  {Ejlli}}\ and\ \bibinfo {author} {\bibfnamefont {V.~R.}\ \bibnamefont
  {Thandlam}},\ }\href {\doibase 10.1103/PhysRevD.99.044022} {\bibfield
  {journal} {\bibinfo  {journal} {Phys. Rev.}\ }\textbf {\bibinfo {volume}
  {D99}},\ \bibinfo {pages} {044022} (\bibinfo {year} {2019})},\ \Eprint
  {http://arxiv.org/abs/1807.00171} {arXiv:1807.00171 [gr-qc]} \BibitemShut
  {NoStop}%
%%CITATION = ARXIV:1807.00171;%%
\bibitem [{\citenamefont {Golat}\ \emph {et~al.}(2019)\citenamefont {Golat},
  \citenamefont {Lim},\ and\ \citenamefont
  {Rodríguez-Fortuno}}]{Golat:2019aap}%
  \BibitemOpen
  \bibfield  {author} {\bibinfo {author} {\bibfnamefont {S.}~\bibnamefont
  {Golat}}, \bibinfo {author} {\bibfnamefont {E.~A.}\ \bibnamefont {Lim}}, \
  and\ \bibinfo {author} {\bibfnamefont {F.~J.}\ \bibnamefont
  {Rodríguez-Fortuno}},\ }\href@noop {} {\  (\bibinfo {year} {2019})},\
  \Eprint {http://arxiv.org/abs/1903.09690} {arXiv:1903.09690 [astro-ph.CO]}
  \BibitemShut {NoStop}%
%%CITATION = ARXIV:1903.09690;%%
\bibitem [{\citenamefont {Morozov}\ and\ \citenamefont
  {Pustovoit}(2020)}]{Morozov2020}%
  \BibitemOpen
  \bibfield  {author} {\bibinfo {author} {\bibfnamefont {A.~N.}\ \bibnamefont
  {Morozov}}\ and\ \bibinfo {author} {\bibfnamefont {V.~I.}\ \bibnamefont
  {Pustovoit}},\ }\href@noop {} {\bibfield  {journal} {\bibinfo  {journal}
  {Herald of the Bauman Moscow State Technicla University, Series Natural
  Sciences}\ }\textbf {\bibinfo {volume} {1(88)}},\ \bibinfo {pages} {46}
  (\bibinfo {year} {2020})}\BibitemShut {NoStop}%
\end{thebibliography}%

\end{document}